\documentclass{article}
\title{Spinor representation of Lie algebra for complete linear group}

\author{C.V. Usenko${}^\dag$  and B.I. Lev${}^\ddag$\\
\\${}^\dag$ \scriptsize{Kyiv National University, Department of Theoretical Physics}
\\${}^\ddag$ \scriptsize{Institute of Physics NAS Ukraine, Department of Theoretical Physics}
}
\date{}
\begin{document}
\maketitle

\abstract{Spinor representation of group GL(4,R) on special spinor space is developed. Representation space has a structure of the fiber space with the space of diagonal metricses as the base and standard spinor space as typical fiber.

Non-isometric motions of the space-time entail spinor transformations which are represented by translation over fibering base in addition to standard $Spin(4,C)$ representation.}

\section{Introduction}

Spinor representation of group GL(4,R) is needed for correct description 
of the Fermi fields on Riemann space, such as the space-time of general 
relativity. It is used for two purposes: to define the connectivity and 
covariant derivative of spinor field and to define the Lie derivative. 
Recent publications \cite{usenko:vlasov,usenko:cohen,usenko:dubois} have 
reminded of this problem.

The important problem to define Fermi fields on Riemann space is that transformation properties of Dirac equation correspond \cite{usenko:wheeler} to $Spin(3,1)$ 
representation of Lorentz group $SO(3,1)$ only, not the full linear group 
$GL(4,R)$. 

Covariant derivative definition and based on it field equations can be defined by the field of orthonormal basis - tetrad description of curve geometry. On such way the tetrad connectivity is a member of Lorentz group and generate the spinor connectivity as standard Spin(3,1) representation.

Spinor representation of group GL(4,R) is needed for investigations of the spinor field symmetry as realization of the space-time symmetry. 

In the case the space-time symmetry subgroup G is different to SO(3,1), the subgroup spinor representation of that symmetry can not be realized as Spin(3,1) subgroup and one needs the spinor representation of G. As example we can take up the standard model of Universe and its G(6) group of symmetry. It contains the subgroup G(3) of isotropy - subgroup of Lorentz group SO(3,1), and subgroup G(3) of translations. Last is not a part of Lorentz group and we can describe translation properties of spinor field (i.e. electron) through spinor representation of group GL(4,R) only.

Here we give results of investigations in special construction for the 
spinor field on the space-time of general relativity. 
This is the extention of our investigation \cite{usenko:self} of spinor representation for full linear group $GL(4,R)$.

\section{Standard construction on Riemann space}

For each point of the space-time one constructs the orthonormal basis 
${\mathop e\limits^k} _\mu \left( x \right)$ such that scalar products are
\begin{equation}\label{usenko:eq1}
{\mathop e\limits^k} _\mu \left( x \right){\mathop e\limits^m} _\nu \left( x 
\right)g^{\mu \nu } = \eta ^{km} = diag\left( {1, - 1, - 1, - 1} \right).
\end{equation}

Each basis vector is represented by Dirac matrix: 
\begin{equation}\label{usenko:eq2}
{\mathop e\limits^k} \Rightarrow \gamma ^k;\gamma ^0 = \left( 
{{\begin{array}{cc}
 1  & 0  \\
 0  &  - 1 \\
\end{array} }} \right),\vec {\gamma } = \left( {{\begin{array}{cc}
 0  & \vec {\sigma } \\
 { - \vec {\sigma }}  & 0  \\
\end{array} }} \right).
\end{equation}

Coordinate transformations deal with the coordinate index only:
\[
{\mathop e\limits^k} _{\mu' } \left( {x'} \right) = \frac{\partial x^\mu 
}{\partial x^{\mu' }}{\mathop e\limits^k} _\mu \left( x \right);
\]
\noindent
and group of invariance for the basis is $SO(3,1)$. This group can be 
represented by transformations of the spinor field: 
\begin{equation}\label{usenko:eq3}
{\mathop {e'}\limits^k} _\mu \left( x \right) = T_m^k {\mathop e\limits^m} _\mu 
\left( x \right) \Rightarrow {\psi' }\left( x \right) = U\left( T 
\right)\psi \left( x \right).
\end{equation}
\begin{equation}\label{usenko:eq4}
U=\exp\left( \frac{i}{4}T_{mn}\sigma^{mn} \right);\;
\sigma ^{nm} = \frac{i}{2}\left( {\gamma 
^n\gamma ^m - \gamma ^m\gamma ^n} \right)
\end{equation}
Main problem in spinor representation of basis transformations is in the 
existence of special type of conjugacy for Dirac spinor: spinor $\overline 
\psi $ is conjugated to $\psi = \left( {{\begin{array}{c}
 \varphi  \\
 \chi  \\
\end{array} }} \right)$ if its components are not conjugated only, but are 
supplementary rearranged by matrix $\gamma^0$: 
\begin{equation}\label{usenko:eq5}
\overline \psi = \left( 
{{\begin{array}{*{20}c}
 {\varphi ^\ast } \hfill & {\chi ^\ast } \hfill \\
\end{array} }} \right)\gamma ^0.
\end{equation}
Each transformation modifying that matrix deforms the norm of the spinor 
space and invariance loses the physical sense.

Covariant derivative can be defined by means of $SO(3,1)$ representation only. 
 The space-time connectivity for basis ${\mathop e\limits^k} _\mu \left({x + dx} \right) $ 
 is defined by
\begin{equation}\label{usenko:eq6}
{\mathop e\limits^k} _\mu \left( 
{x + dx} \right) = {\mathop e\limits^k} _\mu \left( x \right) + dx^\nu \omega 
_{m\nu }^k {\mathop e\limits^m} _\mu \left( x \right)
\end{equation}
\noindent 
This leads to the spinor connectivity 
\begin{equation}\label{usenko:eq7}
\psi \left( {x + dx} \right) = \psi \left( x 
\right) + \frac{i }{ 4}dx^\nu \omega _{m\nu }^k \eta _{kn} \sigma 
^{nm}\psi 
\end{equation}
\noindent and the covariant derivative
\begin{equation}\label{usenko:eq8}
\nabla_\mu\psi \left( x \right)=
\partial_\mu \psi \left( x \right) + \frac{i }{ 4} \omega _{m\mu }^k \eta _{kn} \sigma 
^{nm}\psi
\end{equation}
for spinor field.

This is typical approach to involve Fermi fields in general relativity but 
it is  unsuitable to define the Lie derivative of spinor.
 Until one  considers Lie derivative along Killing vector only, one can 
keep to previous representation. 

If it is needed to involve Lie derivative along non Killing vector one has to use 
a corresponding element of group GL(4,R) being outside the Lorentz group. 
One can need such a Lie derivative, for example, in the case of 
investigation of spinor field time dependence for non static Universe. 

\section{Point-to-point transformation}

We investigate properties of spinor field with respect to motion of the space-time 
\begin{equation}\label{usenko:eq21}
M:x \to y = m\left( 
x \right).
\end{equation}
\noindent Each map $m\left( x \right)$ of this motion 
generates transformation of coordinate basis
\begin{equation}\label{usenko:eq22}
y = m\left( x \right) 
\Rightarrow T_\nu ^\mu = \frac{\partial m^\mu \left( x \right)}{\partial 
x^\nu }.
\end{equation}
 When the motion belongs to neighborhood of identity, this 
transformation takes exponential form
\begin{equation}\label{usenko:eq23}
T = \exp \left( {m \cdot t} \right),	
\end{equation}
where 
\begin{equation}\label{usenko:eq24}
t_\nu ^\mu = \frac{\partial \zeta ^\mu \left( x \right)}{\partial x^\nu }	
\end{equation}
\noindent and vector $\zeta ^\mu \left( x \right)$ 
determines the direction of motion.
 
Derivative of basis along motion is the Lie derivative
\begin{equation}\label{usenko:eq25}
L_\zeta \mathop 
{e_\mu }\limits^k \left( x \right) = \zeta ^\nu \left( x \right)\partial 
_\nu \mathop {e_\mu }\limits^k \left( x \right) + \mathop {e_\nu }\limits^k 
\left( x \right)\partial _\mu \zeta ^\nu \left( x \right).	
\end{equation}
 It generates the transformation of basis 
\begin{equation}\label{usenko:eq26}
\mathop e\limits^k \left( {x + \tau \zeta } \right) = 
\mathop e\limits^k \left( x \right) + \tau L_\zeta \mathop e\limits^k \left( 
x \right),	
\end{equation}
 which can be rewriten as
\begin{equation}\label{usenko:eq27}
\mathop e\limits^k \left( {x + \tau \zeta 
} \right) = \mathop e\limits^k \left( x \right) + \tau \zeta _m^k \mathop 
e\limits^m \left( x \right).	
\end{equation}
 After integrating we obtain basis transformation as   
representation of group $GL(4,R)$
\begin{equation}\label{usenko:eq28}
\mathop e\limits^k \left( {m\left( x \right)} \right) = \exp 
\left( {m\zeta _m^k } \right)\mathop e\limits^m \left( x \right).	
\end{equation}
Only in the case of
$\zeta ^\mu \left( x \right)$ being a Killing vector, this representation 
can be continued to the spinor transformation. In general case this does not work and it is needed to extend the spinor space.

\subsection{Space of diagonal metrics}
A transformation from neighborhood of identity can be represented as a product of 
two isometrics $V$, $U$ and dilatation
\begin{equation}\label{usenko:eq29}
\Delta =  \left( {{\begin{array}{cccc}
 {\delta_0 }  & 0  & 0  & 0  \\
 0  & {\delta_1 }  & 0  & 0  \\
 0  & 0  & {\delta_2 }  & 0  \\
 0  & 0  & 0  & {\delta_3 }  \\
\end{array} }} \right) \quad T = V \cdot \Delta \cdot U.	
\end{equation}
 Both isometrics have  spinor 
representation, but the dilatation has no, because it deforms spinor 
conjugation. One has to extend the spinor space to represent the subgroup of 
dilatation. 

The subgroup of dilatation is a noncompact Abelian group and has true 
representations as translations in $R^{4}$. 

Thus it turnes out to be interesting to 
involve into consideration the space of diagonal metricses $D_m$.
\begin{equation}\label{usenko:eq2a}
D_m=\left\{diag\left(d_0 ,d_1 ,d_2 , d_3 \right):d_0 \cdot d_1 \cdot d_2 \cdot d_3 \neq 0 \right\}	
\end{equation}
 This 
space realizes representation of the dilatation subgroup $\Delta$ which is extended to the representation of group $G(4,R)$ in such a way: 

A point $d$ of $D_m$ is transformed 
by the element of group $g$ to a symmetric matrix $d_g$ which has the diagonal form $d'$. This diagonal matrix determines the reflex of $d$ trough transformation $T_g(d)=d'$ and determines also unique element 
$\Delta_g$ of dilatation subgroup. Left isometrics $V_g$ is exactly the same as $d_g$ transformation 
 to $d'$ and right isometrics $U_g$ can by restored uniquely trough 
\begin{equation}\label{usenko:eq2b}
U_g=\Delta_{g}^{ - 1}V_{g}^{ - 1}T_g.	
\end{equation}
\subsection{Spinor fiber space}

Now we construct for each point $d$ from space of diagonal metricses $D_m$ the spinor space $Spin(4,C)$ with 
anticommutator 
\begin{equation}\label{usenko:eq2c}
\gamma ^n\gamma ^m + \gamma ^m\gamma ^n = 2d^{mn},	
\end{equation}
\[d^{mn} = 
diag\left( {d_0 ,d_1 ,d_2 ,d_3 } \right);\]
\[
\left(\gamma ^0 \right)^2=d_0;
\]
and with conjugation
\begin{equation}\label{usenko:eq2d}
\overline \psi 
= \left( {{\begin{array}{cc}
 {\varphi ^\ast }  & {\chi ^\ast }  \\
\end{array} }} \right)\gamma ^0.	
\end{equation}
Each spinor space $Spin(4,C)$ realize spinor representation of 
isometric group $SO(3,1)$ for metrics $d^{mn}$. All spinor spaces are isomorphous and can be attached to fiber 
space with base $D_m$.

Now non-isometric motion $M:x \to y = m\left( x \right)$, for each point $x$ 
from space-time, which has exponential form (\ref{usenko:eq23})
is represented as product of 
two isometrics 
\begin{equation}\label{usenko:eq2e}
V_g = \exp \left( {v_m \cdot v_g} \right) ;U_g= \exp \left( {u_m \cdot  u_g} \right);	
\end{equation}
and dilatation
\begin{equation}\label{usenko:eq2f}
\Delta_g = \exp \left( {d_m \cdot  \delta_g} \right);	
\end{equation}
as matrix exponent
\begin{equation}\label{usenko:eq2g}
T_g = \exp \left( {v_m \cdot  v_g} \right) \cdot \exp \left( 
{d_m \cdot  \delta_g} \right) \cdot \exp \left( {u_m \cdot  u_g} \right).	
\end{equation}
The motion $T_g$ is represented on fiber spinor space in three 
steps: 
\begin{enumerate}
	\item 
Right isometrics $U_g$ in start point $d$
\begin{equation}\label{usenko:eq2h}
U_g:\psi \left( {x;d} \right) \Rightarrow \exp \left( {u_m \cdot 
u_g\left( x \right)} \right)\psi \left( {x;d} \right);	
\end{equation}
\item
 Translation from start point $d$ to end point $d + \delta_g$ over the base of fiber spinor space and to end point $m(x)$ over space-time
\begin{equation}\label{usenko:eq2i}
\Delta_g:\psi \left( {x;d} 
\right) \Rightarrow \exp \left( {u_m \cdot  u_g\left( x \right)} \right)\psi \left( {x 
;d + \delta_g} \right);	
\end{equation}

\item
  Left isometrics $V_g$ in end point $d +  \delta_g$
\begin{equation}\label{usenko:eq2j}
V_g:\exp \left( {u_m \cdot  u_g \left( x \right)} \right)\psi \left( 
{ x ;d +  \delta_g } \right) \Rightarrow	\exp \left( {v_m \cdot  v_g\left( 
{m\left( x \right)} \right)} \right)\exp \left( {u_m \cdot  u_g\left( x \right)} 
\right)\psi \left( { x ;d + \delta_g } \right)
\end{equation}
\noindent for the translated spinor.
\end{enumerate}

As result we have the representation
\begin{equation}\label{usenko:eq2k}
T:\psi \left( {x;d} \right) \Rightarrow \psi_g \left( {m\left( x \right);d } \right)=
\exp \left( {v_m \cdot v_g\left( {m\left( x \right)} \right)} \right)\exp \left( 
{u_m \cdot u_g\left( x \right)} \right)\psi \left( { x ;d + \delta_g } 
\right),	
\end{equation}
\noindent which preserves the spinor norm, if measure function on 
the fibering base is translationally invariant.

\section{Example: Lie transformation along time direction. }

As an example we consider the Lie transformation along time for standard model 
of Universe. Metrics of space-time for this model can be written as 
\begin{equation}\label{usenko:eq31}
	ds^2 = dt^2 - R^2 \left( t \right)dl^2
\end{equation}
where $dl^2$
 is the metrics of corresponding 
space. Transformation from $t_{1}$ to $t_{2}$ is dilatation with 
matrix
\begin{equation}\label{usenko:eq32}
T=\frac{R\left( {t_2 } \right)}{R\left( {t_1 } \right)}diag\left( 
{0,1,1,1} \right)	
\end{equation}
and is represented simply as translation through base of the spinor fiber space. 
Corresponding Killing vector acts on the spinor field as derivative along 
direction $\left( {0,1,1,1} \right)$ over base $D_{m}$. 
\begin{equation}\label{usenko:eq33}
L_t \psi \left( {d;t,x} \right) = \frac{\partial }{\partial t}\psi \left( 
{d;t,x} \right) + \left( {\frac{\partial }{\partial d_1 } + \frac{\partial 
}{\partial d_2 } + \frac{\partial }{\partial d_3 }} \right)\psi \left( 
{d;t,x} \right)
\end{equation}

\section{Conclusion}
\begin{itemize}
	\item 
We have developed the special spinor space which represents the full linear group $GL(4,R)$.
It has a structure of the fiber space with the space of diagonal metricses as the base and standard spinor space as typical fiber.

	\item 
Non-isometric motions of the space-time entail spinor transformations which are represented by  translation over fibering base in addition to standard $Spin(4,C)$ representation.

	\item 

Until we do not use the non-isometric motion, spinor fields without 
overlapping on the base of spinor fiber space are independent. Moreover, each such field can 
be represented as simple spinor space
\begin{equation}\label{usenko:eq41}
\psi \left( x \right) \Rightarrow \int {\psi 
\left( {x;d} \right)d^4d}.	
\end{equation}

	\item 
 Only if one  uses the non-isometric motion of 
space-time, it is essential to consider the fibering of the spinor space.
\end{itemize}

\end{document}